# COLLECTIVE INTELLIGENCE IN HUMANS: A LITERATURE REVIEW

Juho Salminen

Lappeenranta University of Technology, Lahti School of Innovation
Saimaankatu 11
15150 Lahti, Finland
e-mail: juho.salminen@lut.fi

## ABSTRACT

This literature review focuses on collective intelligence in humans. A keyword search was performed on the Web of Knowledge and selected papers were reviewed in order to reveal themes relevant to collective intelligence. Three levels of abstraction were identified in discussion about the phenomenon: the micro-level, the macro-level and the level of emergence. Recurring themes in the literature were categorized under the above-mentioned framework and directions for future research were identified.

## INTRODUCTION

Study of collective intelligence in humans is a relatively new field, for which huge expectations are set, for example through speculations on the emergence of the Global Brain (see Heylighen, in press, for a review). Especially new forms of collaboration made possible by the Internet, web 2.0 and social media add to the hype. It is, therefore, no wonder that interest in the field is rising.

According to an often-cited definition, collective intelligence is a form of universal, distributed intelligence, which arises from the collaboration and competition of many individuals (Levy 1997). It is the general ability of a group to perform a wide variety of tasks (Woolley et al. 2010). The phenomenon is closely related to swarm intelligence, which means collective, largely self-organized behavior emerging from swarms of social insects (Bonabeau and Meyer 2001). These terms have been used somewhat interchangeably; for example, Krause et al. (2009) define swarm intelligence as "two or more individuals independently, or at least partially independently, acquire information and these different packages of information are combined and processed through social interaction, which provides a solution to a cognitive problem in a way that cannot be implemented by isolated individuals". For the remainder of the paper, I will use swarm intelligence to refer to the emergent, collective behavior of groups of cognitively simple agents such as insects, robots and simulation algorithms. The term collective intelligence is reserved for phenomena involving agents with high cognitive capabilities, namely humans. This distinction is in line with the use of terminology in the literature. About 25 % of the papers found on Web of Knowledge using keywords 'collective intelligence' discuss humans, while only 2 % of the papers found with keywords 'swarm intelligence' do so.

Approaches to studying collective intelligence have been diverse, from the purely theoretical (Szuba 1998; 2002) and conceptual (Luo et al. 2009) to simulations (Bosse et al. 2006), case studies (Gruber 2007), experiments (Woolley et al. 2010) and systems design (Vanderhaeghen and Vettke 2010). The field is also multidisciplinary as it is related to, at least, psychology (Woodley and Bell 2011), complexity sciences (Schut 2010), cognitive studies (Trianni et al. 2011), biology (Bonabeau and Meyer 2001), computer sciences and semantics (Levy 2010) and social media (Shimazu and Koike 2007). At the moment, there is no theory capable of explaining how collective intelligence actually works (Schut 2010). Despite some efforts (e.g. Luo et al. 2009, Gan et al. 2007, Malone et al. 2010), generally accepted frameworks for studying collective intelligence in humans do not exist either, and as a result, the field might be at risk of fragmentation. Although a certain amount of diversity is probably good for the advancement of a scientific field (Woolley and Fuchs 2011), a lack of overarching structure could make the field appear confusing and make it challenging to tie the efforts of different disciplines together in a coherent way. Furthermore, due to the lack of a common framework, it is not possible to assess what is already known. It is challenging for researchers from different disciplines to be aware of advancements in other fields, possibly under differently named concepts.

This paper focuses on the question of what scientific community means by the notion of collective intelligence in human context. The objective is to review current literature, identify relevant themes and form a conceptual framework for studying the phenomenon. The scope of the review is limited to literature discussing collective intelligence in humans. The limitation is based on the assumption that rich and complex cognitive and psychological behavior sets humans apart from insects, algorithms and robots. Furthermore, the development of human intelligence resulted in part from evolutionary pressures to navigate in social situations to one's own benefit (Geary 2005). Such behavior could plausibly undermine the collective performance of groups, or at least make it significantly different from situations where motivations of individuals are mostly aligned. At this point it would be premature to try to combine phenomena from different contexts under one framework without first understanding each context separately.

The literature review reveals three levels of abstraction in the discussion about collective intelligence in humans: the micro-level, the macro-level and the level of emergence. This conceptual framework is used to organize relevant themes and to identify directions for further research.

## **METHODS**

The selection of literature for this review follows the approach of Zott et al. (2011). A keyword search was conducted on the Web of Knowledge on 7 July 2011 using the keywords 'collective intelligence' and 'swarm intelligence'. The searches produced 405 and 646 results, respectively. In addition, all issues of the journals Swarm Intelligence and the International Journal of Swarm Intelligence were reviewed for suitable articles.

A cursory analysis was performed by reading through the titles and abstracts. The following criteria were used to select the papers for review: 1) the paper discusses collective intelligence in human context; 2) the publication in which the paper is published is listed on the Web of Knowledge and 3) the paper makes a non-trivial contribution to the discussion about collective intelligence (i.e. it involves more than a couple of mentions of the term).

Using these criteria, 41 papers were selected. The papers are marked with an asterisk (*) in the reference list. The purpose was not to cover everything that has been written about the topic, but to review a representative sample of papers to gain a sufficient understanding of the relevant themes of collective intelligence on humans. The papers were read thoroughly and definitions of collective intelligence and related terms, themes discussed and the main contributions to collective intelligence research were identified. Similar definitions and themes were grouped together and the resulting categories were named as seemed appropriate. Sticky notes were used to make the process visual and thus help the recognition of interesting patterns in the data. Additional references were gathered and further limited literature searches were performed to fill in the gaps (e.g. definitions of self-organization, trust and emotional intelligence) and thus to provide a more complete view on what is already known about collective intelligence in humans.

## **RESULTS**

The grouping of themes and definitions revealed a pattern in the literature. The discussion about collective intelligence in humans appears to be divided into three levels of abstraction: micro-level, macro-level and level of emergence. In Table 1, the themes identified in the literature are grouped under the three levels of abstraction and examples of papers discussing these themes are given. Next, the characteristics of each level are discussed in more detail.

### **The micro-level: Enabling factors of human beings**

At the micro-level, collective intelligence is a combination of psychological, cognitive and behavioral elements. Pentland (2007) argues that humans should firstly be viewed as social animals and only secondarily as individuals. According to his research with the so-called Socioscope, human behavior is largely predictable, non-linguistic signal-response behavior. The immersion of self in a social network is a typical human condition and our unconscious ability to read and display social signals allows smooth coordination within the network. Pentland suggests that important parts of human intelligence could thus reside in network properties. This might just be the case, as Woolley et al. (2010) found evidence of the existence of a single dominant collective intelligence factor, c, underlying group performance. In their experiments, c explained 30-40 % of group performance and was found to depend on the composition of the group (e.g. average intelligence) and emergent factors resulting from interaction of group members, such as conversational turn-taking. Furthermore, c is positively correlated with social sensitivity and the proportion of females in the group, but the influence of females is probably mediated by their better average social sensitivity (Woolley et al. 2010). Many open questions remain regarding the nature of c. Woodley and Bell (2011) suggest that c could actually be largely a manifestation of the General Factor of Personality (Just 2011) at a group level.

Other relevant themes are trust (Scarlat and Maries 2009, Bosse et al. 2006) and attention (Zembylas and Vrasidas 2005, Gruber 2007, Trianni et al. 2011). A certain level of trust is a precondition for cooperation. Attention is used as an implicit measurement of value in many contemporary web applications, such as YouTube (view count) and Twitter (re-tweets).

| Level | Theme | Definition | Examples of papers from the sample |
|---|---|---|---|
| Micro | Humans as social animals | Viewing humans as social animals: immersion of self in a social network a typical human condition | Pentland 2006, Pentland 2007 |
| | Intelligence | The intelligence of individual human beings, often measured with the g-factor | Woolley et al. 2010 |
| | Personal interaction capabilities | The factors affecting a person's ability to interact with other human beings, such as emotional intelligence (Cherniss 2010), social sensitivity (Woolley et al. 2010) and the general factor of personality (Just 2011) | Woolley et al. 2010, Woodley and Bell 2011 |
| | Trust | An actor's expectation of the other party's competence and goodwill (Blomqvist 1997) | Bosse et al. 2006, Scarlat and Maries 2009 |
| | Motivation | The factors influencing the interest to participate in communities or to contribute to collective effort | Franck 2002, Rasmussen et al. 2003, Bonabeau 2009, Lykourentzou et al. 2010, Brabham 2010, Malone et al 2010 |
| | Attention | The commitment of cognitive resources | Zembylas and Vrasidas 2005, Zettsu and Kiyoki 2006, Gruber 2007, Trianni et al. 2011 |
| | Communities | Real and virtual communities, such as communities of practice and online social networks (Cachia et al. 2007) and brand communities (Brabham 2010) | Coe et al. 2001, Cachia et al. 2007, Chen 2007, Lykourentzou et al. 2010, Brabham 2010 |
| Emergence | Complex adaptive systems | Systems that show adaptivity, self-organization and emergence (Ottino 2004) | Komninos 2004, Chen 2007, Luo et al. 2009, Schut 2010, Trianni et al. 2011, |
| | Self-organization | The emergence of order at the system level without central control, solely due to local interactions of the system's components (Kauffman 1993) | Bonabeau and Meyer 2001, Franck 2002, Rasmussen et al. 2003, Wu and Aberer 2003, Luo et al. 2009, Krause et al. 2009, Schut 2010, Trianni et al. 2011 |
| | Emergence | A rise of system level properties that are not present in its components; "the whole is more than the sum of its parts" (Damper 2000) | Rasmussen et al. 2003, Chen 2007, Cachia et al. 2007, Luo et al. 2009, Schut 2010, Lee and Chang 2010, Woolley et al. 2010, Trianni et al. 2011, |
| | Swarm intelligence | The study of cognitively (relatively) simple entities, whose collective behavior is intelligent | Bonabeau and Meyer 2001, Wu and Aberer 2003, Krause et al. 2009, Luo et al. 2009, Trianni et al. 2011, |
| | Stigmergy | A mechanism of indirect coordination, originally describing the nest-building behavior of termites (Theraulaz and Bonabeau 1999) | Bosse et al. 2006 |
| | Distributed memory | The shared, often external, dynamic memory system that performs parts of agents' cognitive processes (Bosse et al 2006) | Bosse et al. 2006, Scarlat and Maries 2009, Gregg 2009, Luo et al. 2009, Levy 2010, Trianni et al. 2011 |
| Macro | Decision making | The process of making decisions, both individually and in groups | Pentland 2006, Bonabeau 2009, Malone et al. 2010, Gregg 2010, Krause et al. 2011 |
| | Wisdom of crowds | Under certain conditions, groups can be more intelligent than the smartest individuals in them; a collective estimate can be accurate, even if individual estimations are not (Surowiecki 2005) | Chen 2007, Pentland 2007, Nguyen 2008, Krause et al. 2009, Brabham 2009, Lykourentzou et al. 2010, Leimeister 2010, Lee and Chang 2010, Brabham 2010, Lorenz et al. 2011, |
| | Aggregation | The combination of individual pieces of information to form a synthesis or collective estimation | Pentland 2007, Bothos et al. 2010, Krause et al. 2011, |
| | Bias | The tendency of individuals and groups to make systematical errors in decision making situations | Cachia et al. 2007, Gregg 2009, Lee and Chang 2010, Krause et al. 2011 |
| | Diversity | The differences in demographic, educational and cultural backgrounds and the ways that people represent and solve problems (Hong and Page 2004) | Bonabeau and Meyer 2001, Bonabeau 2009, Brabham 2010, Krause et al. 2011 |
| | Independence | The decision of an individual is not influenced by the decisions of other individuals | Lorenz et al. 2011 |

*Table 1: A list of themes related to collective intelligence in humans categorized under three levels of abstraction.*

**The macro-level: Output of the System**

At the macro-level, collective intelligence becomes a statistical phenomenon, at least in the case of the 'wisdom of crowds' effect (Lorenz et al. 2011). The term 'wisdom of crowds' was coined by Surowiecki (2005) and it describes a phenomenon where, under certain conditions, large groups can achieve better results than any single individual in the group. For example, the average of several individuals' estimates can be accurate even if individual estimations are not. The 'wisdom of crowds' effect is claimed to be based on diversity, independence and aggregation (Surowiecki 2005).

Diversity in groups of people usually refers to differences in demographic, educational and cultural backgrounds and differences in the ways that people represent and solve problems (Hong and Page 2004). Both a simulation model (Hong and Page 2004) and an experiment with humans (Krause et al. 2011) have shown that under certain conditions groups of diverse problem solvers can outperform groups of high-ability problem solvers. Furthermore, the best problem solvers were biased in their estimations[1], while the group, as a whole, was accurate (Krause et al. 2011). A question remains whether this finding was unusual or something that can be expected in general.

Independence means that the estimations of one individual are not influenced by the estimations of other individuals. Lorenz et al. (2011) have shown that even minor social interaction can undermine the wisdom of crowds, which happens through three effects. The social influence effect reduces the diversity of a group without increasing its accuracy. The range reduction effect causes the correct value to become less central in the distribution of evaluations, thus delivering a false hint regarding the location of the truth. The confidence effect is a psychological result of the two aforementioned statistical effects, and it increases individuals' confidence in their estimations even though collective accuracy has not improved. Lorentz et al. (2011) propose that these effects occur especially in a certain range of difficulty of decision-making and confidence of decision makers. This conjecture should be explored in more detail.

Aggregation refers to mechanisms for pooling and processing individual estimations to a collective estimation. While simple averaging might be the most common method of aggregation, it is not always the most suitable one. In many cases, other statistical aggregate measures should be considered (Lorenz et al. 2011). The rise of the Internet has also made it possible to develop new aggregation methods, such as information aggregation or prediction markets (Bothos et al. 2009), social tagging or folksonomies (Gruber 2007, Zettsu and Kiyoki 2006) and data visualization (Chen 2007).

**The Level of Emergence: From Local Interactions to Global Patterns**

The level of emergence resides between the micro-level and the macro-level and deals with the question of how system behavior on the macro-level emerges from interactions of individuals at the micro-level. A common approach to explaining how collective intelligence as a statistical or probabilistic phenomenon emerges from individual interactions is to use the theories of complex adaptive systems. Complex (adaptive) systems are characterized by adaptivity, self-organization and emergence (Ottino 2004). Adaptivity means the ability of a system, or its components, to change themselves according to changes in the environment (Schut 2010). Self-organization means the emergence of order at the system level without central control, solely due to local interactions of the system's components. The basic ingredients of self-organization are positive and negative feedback loops, randomness and multiple interactions (Bonabeau 1999). A simple definition for emergence is "the whole is more than the sum of its parts" (Damper 2000). Extending from these premises, Schut (2010) proposes three enabling properties and five defining properties for collective intelligence systems. The existence of adaptivity, interaction and rules executed at a local level make it possible for collective intelligence to emerge from a system. If the system can be observed to show a distinction between global and local, randomness, emergence, redundancy and robustness, the system is a collective intelligence one.

Group memory (Trianni et al. 2011), a shared extended mind (Bosse et al. 2006) and other similar concepts are also relevant to the emergence of collective intelligence. Bosse et al. (2006) give the following criteria for a shared extended mind:
- The environment participates in the agents' mental processes.
- The agents' internal mental processes are simplified.
- The agents have a more intensive interaction with the world.
- The agents depend on the external world in the sense that they delegate some of their mental representations and capabilities to it.

A shared extended mind thus works as a dynamic short-term memory that allows the coordination and collaboration of individual components of the complex adaptive system. Notably, the components creating the shared extended mind need not be aware of it, nor benefit from its creation (Bosse et al. 2006).

---

[1] Page (2007) has also proposed that best problem solvers could be systematically biased.

The literature provides plenty of examples of swarm intelligence systems that display the characteristics of complex adaptive systems and a shared extended mind. Behavior of social insects is maybe the most classical example. The foraging of ants (Camazine et al. 2001), the nest-site selection of honeybees (Seeley and Buhrman 1999) and the nest building of termites (Turner 2011) all use some form of distributed memory and show emergent, adaptive behavior as a result of self-organization. The features of complex adaptive systems have also been considered to be relevant in the context of human collective intelligence (Komninos 2004, Luo et al. 2009, Chen 2007) and at least some of the features have been demonstrated in case studies (Wu and Aberer, 2003, Bonabeau 2009, Lykourentzou et al. 2010). The Internet as a shared memory of humankind has been mentioned repeatedly (e.g. Levy 2010, Luo et al. 2009, Heylighen 1999).

**DISCUSSION AND CONCLUSIONS**

A sample of literature discussing the collective intelligence in humans was reviewed and the discovered themes were categorized into micro-level, macro-level and emergence-level phenomena. The framework is similar to the conceptual model of Luo et al. (2009), the gist of which is the question of how macro-level phenomena emerge from micro-level interactions. The framework proposed in this paper emerged from data collected from contemporary literature. Therefore, it is arguable that the scientific community has already implicitly divided collective intelligence to the aforementioned three levels of abstraction. Making this division explicit hopefully brings some structure to the discussion and helps in fitting the pieces of the puzzle together. The categorization of themes related to collective intelligence (Table 1) provides guidance for selecting topics for further literature reviews and suggests how the results might fit into the big picture of collective intelligence in humans.

Based on the framework, I propose that 1) the micro-level features of human beings, such as intelligence, trust and motivation, are the enabling factors of collective intelligence. They provide the 'rules' according to which individuals act. Micro-level features set humans apart from other collective intelligence systems; for example, motivation does not have to be taken into account when designing robots or algorithms. 2) Individuals interacting with each other form a complex adaptive system, which shows self-organization and emergence. Distributed memory facilitates communication and coordination between individuals. A comparison of collective intelligence in humans to examples of swarm intelligence in other contexts might be most fruitful at this level of abstraction. Finally, 3) the global behavior of the complex adaptive system is probabilistic by nature. At this level diversity, independence and mechanisms of information aggregation are important features of the system. Measuring them could be useful in predicting the global performance of the system as a whole.

The framework allows some educated guesses on how other phenomena might be connected to collective intelligence. For instance, promising results have been obtained from using theatre-based methods in relieving organizational issues (Pässilä and Oikarinen 2011). As "improvisation theatre is about interaction" (Minna Partanen, personal communication), it can be hypothesized that theatre-based methods contribute to collective intelligence by influencing human interaction at the micro level. Visualization tools for group work, such as sticky notes and shared visual templates (e.g. Sibbet 2010), could be interpreted as shared, dynamic memory systems which facilitate the functioning of complex adaptive systems. Finally, using Twitter searches to monitor discussions on social media arguably increases the probability of stumbling upon some relevant new information. Complex interactions of millions of users manifest themselves as a probabilistic phenomenon in a way that has even been compared to the workings of a brain (Pomerlau 2009)[2]

The proposed framework points out some directions for future research. There is a built-in hypothesis that at the macro-level collective intelligence is a statistical or probabilistic phenomenon. The 'wisdom of crowds' effect and the collective intelligence factor c have this characteristic, but how about other examples of collective intelligence? Many researchers have pointed out the lack of sufficient theory on collective intelligence (Bonabeau 2009, Schut 2010, Luo et al. 2009) and it is most notable at the level of emergence. A better understanding is needed on how micro-level activities lead to macro-level behavior in human contexts. This requires a multidisciplinary approach and simulations to identify the underlying mechanisms of cognitive processes (Trianni et al. 2011). Schut (2010) provides a good overview on the design of models for simulating collective intelligence, and previous research for example on group performance (Kerr and Tindale 2004) could probably provide directions for efforts. The resulting deeper understanding of cognition as an emergent phenomenon could be used to improve conceptual models and the design frameworks of collective intelligence. These frameworks should also be tested and validated more rigorously. One interesting direction would be to investigate how systems react to violations of factors facilitating collective or swarm intelligence (Krause et al. 2009). Benchmarking social insects in the design of human collective intelligence systems might also be a fruitful direction for future research. For example, stigmergy as a coordination mechanism has shown some promise (Besten et al. 2008).

---

[2] Stumbled upon in a Twitter search. See also Passino et al. (2008) for a related, more academic comparison of honeybee swarms and neurons.

As always, this study has its limitations. The initial sample of literature was obtained from a single database with only two keyword searches. The scope was limited only to papers discussing collective intelligence of humans. This work could be expanded by reviewing literature from other sources and by including also non-human examples. The possibility of mistakes made by the researcher cannot be ruled out. Despite the attempt for scientific rigor, important sources may have been missed during the cursory analysis of the initial sample, and the identification of the themes and their categorization is subjective. The role of the Internet in collective intelligence has been touched on only casually, although it was frequently mentioned in the reviewed literature. However, here the focus was on the principles of collective intelligence, which apply regardless of the existence of the Internet. Although the Internet is a great environment for facilitating collective intelligence, it is not needed for understanding the phenomenon in general.

In conclusion, combining various approaches of studying the collective intelligence of humans seems possible despite the multidisciplinary nature of the phenomenon. The three levels of abstraction offer different lenses through which collective intelligence can be viewed. The viewpoints complement each other to provide a fuller picture of this interesting phenomenon.

## ACKNOWLEDGEMENTS

The author is grateful for anonymous reviewers for their helpful comments and Eelko Huizingh for advices on structuring the paper. The author also wishes to thank the European Regional Development Fund and the Regional Council of Päijät-Häme for the opportunity of presenting his research at the CI2012 conference.

## REFERENCES

Besten, M., Dalle, J-M. and Galia, F. (2008) "The allocation of collaborative efforts in open-source software", *Information Economics and Policy,* **20,** 316-322.

Bonabeau, E. (1999) *Swarm Intelligence: From Natural to Artificial Systems,* Oxford University Press, 305 pages.

*Bonabeau, E. (2009) "Decisions 2.0: The Power of Collective Intelligence," *MIT Sloan Management Review,* **50,** 2: 45-52.

*Bonabeau, E. and Meyer, C. (2001) "Swarm Intelligence: A Whole New Way to Think About Business," *Harvard Business Review,* **79,** 5: 106-114.

*Bosse, T., Jonker, C. M., Schut, M. C. and Treur, J. (2006) "Collective Representational Content for Shared Extended Mind," *Cognitive Systems Research,* **7,** 151-174.

*Bothos, E., Apostolou, D. and Mentzas, G. (2009) "Collective Intelligence for Idea Management with Internet-based Information Aggregation Markets," *Internet Research,* **19,** 26-41.

Blomqvist, K. (1997) "The Many Faces of Trust", *Scandinavian Journal of Management,* **13,** 271-286.

*Brabham, D. C. (2009) "Crowdsourcing the Public Participation Process for Planning Projects," *Planning Theory,* **8,** 242-262.

*Brabham, D. C. (2010) "Moving the Crowd at Threadless," *Information, Communication & Society,* **13,** 1122-1145.

*Cachia, R., Compañó, R. and Da Costa, O. (2007) "Grasping the Potential of Online Social Networks for Foresight," *Technological Forecasting & Social Change,* **74,** 1179-1203.

Camazine, S., Deneubourg, J-L., Franks, N. R., Sneyd, J., Theraulaz, G and Bonabeau, E. (2001) *Self-organization in Biological Systems,* Princeton University Press, 560 pages.

*Chen, C. (2007) "Holistic Sense-making: Conflicting Opinions, Creative Ideas, and Collective Intelligence," *Library Hi Tech,* **25,** 311-327.

Cherniss, C. (2010) "Emotional Intelligence: Toward Clarification of a Concept," *Industrial and Organizational Psychology-Perspectives on Science and Practice,* **3,** 110-126.

*Coe, A., Paquet, G. and Roy, J. (2001) "E-governance and Smart Communities: A Social Learning Challenge," *Social Science Computer Review,* **19,** 80-93.

Damper, R. I. (2000) "Emergence and Levels of Abstraction," *International Journal of Systems Science,* **31,** 811-818.

*Fathianathan, M., Panchal, J. H. and Nee, A. Y. C. (2009) "A Platform for Facilitating Mass Collaborative Product Realization," *CIRP Annals – Manufacturing Technology,* **58,** 127-130.

*Franck, G. (2002) "The Scientific Economy of Attention: A Novel Approach to Collective Rationality of Science," *Scientometrics,* **55,** 3-26.

*Gan, Y. and Zhu, Z. (2007) "A Learning Framework for Knowledge Building and Collective Wisdom Advancement in Virtual Learning Communities," *Educational Technology & Society,* **10,** 206-226.

Geary, D. (2005) *Origin of Mind: Evolution of brain, cognition and general intelligence,* American Psychological Association, 459 pages.

*Gregg, D. (2009) "Developing a Collective Intelligence Application for Special Education," *Decision Support Systems,* **47,** 455-465.

*Gregg, D. (2010) "Designing for Collective Intelligence," *Communications of the ACM,* **53,** 134-138.

*Gruber, T. (2008) "Collective Knowledge Systems: Where the Social Web Meets the Semantic Web," *Journal of Web Semantics,* **6,** 4-13.

Heylighen, F. (1999) "Collective Intelligence and its Implementation on the Web: Algorithms to Develop a Collective Mental Map," *Computational & Mathematical Organization Theory,* **5,** 253-280.

Heylighen, F. (in press) "Conceptions of a Global Brain: an Historical Review," *Technological Forecasting and Social Change,* Retrieved 14 August 2011 from http://pespmc1.vub.ac.be/Papers/GBconceptions.pdf

Hong, L. and Page, S. (2004) "Groups of Diverse Problem-solvers Can Outperform Groups of High-ability Problem-solvers," *PNAS,* **101,** 16385-16389.

Howe, J. (2008) *Crowdsourcing: Why the Power of the Crowd is Driving the Future of Business,* Random House Books, 312 pages.

*Jung, Y., Park, Y., Bae, H. J. and Lee, B. S. (2011) "Employing Collective Intelligence for User Driven Service Creation," *IEEE Communications Magazine,* **49,** 75-83.

Just, C. (2011) "A Review of Literature on General Factor of Personality," *Personality and Individual Differences,* **50,** 765-771.

Kauffman, S. (1993) *The Origins of Order: Self-organization and selection in evolution*, Oxford University Press, 728 pages.

Kerr, N. L. and Tindale, R. S. (2004) "Group Performance and Decision Making," *Annual Review of Psychology,* **55,** 623-655.

*Komninos, N. (2004) "Regional Intelligence: Distributed Localised Information Systems for Innovation and Development," *International Journal of Technology Management,* **28,** 483-506.

*Krause, J., Ruxton, G. and Krause, S. (2009) "Swarm Intelligence in Animals and Humans," *Trends in Ecology and Evolution,* **25,** 28-34.

*Krause, S., James, R., Faria, J. J., Ruxton, G. D. and Krause, J. (2011) "Swarm Intelligence in Humans: Diversity Can Trump Ability," *Animal Behaviour,* **81,** 941-948.

*Lee, J-H. and Chang, M-L. (2010) "Stimulating Designers' Creativity Based on a Creative Evolutionary System and Collective Intelligence in Product Design," *International Journal of Industrial Ergonomics,* **40,** 295-305.

*Leimeister, J. M. (2010) "Collective Intelligence," *Business & Information Systems Engineering,* **2,** 245-248.

Levy, P. (1997) *Collective Intelligence: Mankind's Emerging World in Cyberspace,* Basic Books, 277 pages.

*Levy, P. (2010) "From Social Computing to Reflexive Collective Intelligence: The IEML Research Program," *Information Sciences,* **180,** 71-94.

*Lorenz, J., Rauhut, H., Schweitzer, F. and Helbing, D. (2011) "How Social Influence Can Undermine the Wisdom of Crowd Effect," *PNAS,* **108,** 9020-9025.

*Luo, S., Xia, H., Yoshida, T. and Wang, Z. (2009) "Toward Collective Intelligence of Online Communities: A Primitive Conceptual Model," *Journal of Systems Science and Systems Engineering,* **18,** 2: 203-221.

*Lykourentzou, I., Papadaki, K., Verkados, D. J., Polemi, D. and Loumos V. (2010) "CorpWiki: A Self-regulating Wiki to Promote Corporate Collective Intelligence Through Expert Peer Matching," *Information Sciences,* **180,** 18-38.

*Malone, T. W., Laubacher, R. and Dellarocas, C. (2010) "The Collective Intelligence Genome," *MIT Sloan Management Review,* **51,** 3: 21-31.

*Nguyen, N. T. (2008) "Inconsistency of Knowledge and Collective Intelligence," *Cybernetics and Systems,* **39,** 542-562.

Ottino, J. M. (2004) "Engineering Complex systems," *Nature,* **427,** 339.

Page, S. (2007) *The Difference: How the Power of Diversity Creates Better Groups, Firms, Schools*

*and Societies.* Princeton University Press, 448 pages.

Passino, K.M., Seeley, T. D. and Visscher, P. K. (2008) "Swarm Cognition in Honey Bees," *Behavioral Ecology and Sociobiology,* **62,** 401-414.

*Pentland, A. (2006) "Collective Intelligence," *IEEE Computational Intelligence Magazine*, **1**, 3: 9-12.

*Pentland, A. (2007) "On the Collective Nature of Human Intelligence," *Adaptive behavior,* **15,** 2: 189-198.

Pomerlau, D. (2009) "Twitter and the Global Brain", Retrieved 1 June 2011 from https://docs.google.com/Doc?docid=0AT9namoRlUKMZGM3eG40ZHNfN2Z3ajJ0eGNw&hl=en&pli=1

Pässilä, A. and Oikarinen, T. (2011) "Research-based theatre as a facilitator for organisational learning". In Meusburger, A.P., Berthoin-Antal, A. and Wunder, E. (eds.) *Knowledge in organizations - Learning organizations.* Vol. 6 of the series Learning, Knowledge and Space. Springer.

*Rasmussen, S., Raven, M. J., Keating, G. N. and Bedau, M. A. (2003) "Collective Intelligence of the Artificial Life Community on Its Own Successes, Failures, and Future," *Artificial Life,* **9,** 207-235.

*Scarlat, E. and Maries, I. (2009) "Increasing Collective Intelligence within Organizations Based on Trust and Reputation Models," *Economic Computation and Economic Cybernetics Studies and Research,* **43,** 2: 61-72.

*Schut, M. C. (2010) "On Model Design for Simulation of Collective Intelligence," *Information Sciences,* **180,** 132-155.

Seeley, T. D. and Buhrman, S. C. (1999) "Group Decision Making in Swarms of Honey Bees," *Behavioral Ecology and Sociobiology,* **45,** 19-31.

*Shimazu, H. and Koike, S. (2007) "KM 2.0: Business Knowledge Sharing in the Web 2.0 age," *NEC Technical Journal, 2,* 2: 50-54.

Sibbet, D. (2010) *Visual Meetings: How Graphics, Sticky Notes and Idea Mapping Can Transform Group Productivity,* Wiley, 288 pages.

Surowiecki, J. (2005) *Wisdom of Crowds,* Anchor Books, 306 pages.

*Szuba, T. (2001) "A Formal Definition of the Phenomenon of Collective Intelligence and Its IQ Measure," *Future Generation Computer Systems,* **17,** 489-500.

*Szuba, T. (2002) "Universal Formal Model of Collective Intelligence and Its IQ Measure," *Lecture Notes in Artificial Intelligence,* **2296,** 303-312.

Theraulaz, G. and Bonabeau, E. (1999) "A Brief History of Stigmergy", *Artificial Life,* **5,** 97-116.

*Trianni, V., Tuci, E., Passino, K. M. and Marshall, J. A. R. (2011) "Swarm Cognition: an Interdisciplinary Approach to the Study of Self-organizing Biological Collectives," *Swarm Intelligence,* **5,** 3-18.

Turner, J. S. (2011) "Termites as Models of Swarm Cognition," *Swarm Intelligence,* **5,** 19-43.

*Vanderhaeghen, D. and Fettke, P. (2010) "Organizational and Technological Options for Business Process Management from the Perspective of Web 2.0: Results of a Design Oriented Research Approach with Particular Consideration of Self-Organization and Collective Intelligence," *Business & Information Systems Engineering,* **2**, 15-28.

*Woodley M. A., and Bell, E. (2011) "Is Collective Intelligence (mostly) the General Factor of Personality? A Comment on Woolley, Chabris, Pentland, Hashmi and Malone (2010), *Intelligence,* **39,** 79-81

*Woolley, A. W., Chabris, C. F., Pentland, A., Hashmi, N. and Malone, T. (2010), "Evidence for a Collective Intelligence Factor in the Performance of Human Groups," *Science,* **330,** 686-688.

Woolley, A. W. and Fuchs, E. (2011) "Collective Intelligence in the Organization of the Science," *Organization Science,* **22,** 1359-1367.

*Wu, J. and Aberer, K. (2003) "Swarm Intelligent Surfing in the Web," *Lecture Notes in Computer Science,* **2722,** 431-440.

*Zembylas, M. and Vrasidas, C. (2005) "Globalization, Information, and Communication Technologies, and the Prospect of a 'Global Village': Promises of Inclusion or Electronic Colonization?" *Journal of Curriculum Studies, 2005,* **37,** 1: 65-83.

*Zettsu, K. and Kiyoki, Y. (2006) "Towards knowledge management based on harnessing collective intelligence on the web," *Lecture Notes in Artificial Intelligence,* **4248,** 350-357.

Zott, C., Amit, R. and Massa, L. (2011) "The business model: Recent developments and future research," *Journal of Management,* **37,** 1019-1042.